\documentclass[fleqn,usenatbib]{mnras}

\usepackage{newtxtext,newtxmath}
\usepackage[T1]{fontenc}
\usepackage{ae,aecompl}
\usepackage{tabularx}

\usepackage{graphicx}	
\usepackage{amsmath}	
\usepackage{amssymb}	

\usepackage{BibDef}

\newcommand{\zt}{\tilde{Z}}

\title[The CII--SFR in dwarf galaxies]{The [CII]--SFR correlation in dwarf galaxies across cosmic time}

\author[A. Lupi and S. Bovino]{
Alessandro Lupi$^{1}$\thanks{E-mail: alessandro.lupi@sns.it} and
Stefano Bovino$^{2}$\\
$^{1}$Scuola Normale Superiore, Piazza dei Cavalieri 7, Pisa, IT-56126 Italy\\
$^{2}$Departamento de Astronom\'ia, Universidad de Concepci\'on, Barrio Universitario, Concepci\'on, Chile\\
}

\date{Accepted XXX. Received YYY; in original form ZZZ}

\pubyear{2019}

\begin{document}
\label{firstpage}
\pagerange{\pageref{firstpage}--\pageref{lastpage}}
\maketitle

\begin{abstract}
Current galaxy observations suggest that a roughly linear correlation exists between the [CII] emission and the star formation rate, either as spatially-resolved or integrated quantities. Observationally, this correlation seems to be independent of metallicity, but the very large scatter does not allow to properly assess whether this is true. On the other hand, theoretical models tend to suggest a metallicity dependence of the correlation. In this study, we investigate the metallicity evolution of the correlation via a high-resolution zoom-in cosmological simulation of a dwarf galaxy employing state-of-the-art sub-grid modelling for gas cooling, star formation, and stellar feedback, and that self-consistently evolves the abundances of metal elements out of equilibrium. Our results suggest that the correlation should evolve with metallicity, in agreement with theoretical predictions, but also that this evolution can be hardly detected in observations, because of the large scatter. We also find that most of the [CII] emission is associated with neutral gas at low-intermediate densities, whereas the highest emissivity is produced by the densest regions around star-forming regions.
\end{abstract}


\def\lsim{\mathrel{\rlap{\lower 3pt \hbox{$\sim$}} \raise 2.0pt \hbox{$<$}}}
\def\gsim{\mathrel{\rlap{\lower 3pt \hbox{$\sim$}} \raise 2.0pt \hbox{$>$}}}
\def\msun{\rm {M_\odot}}
\def\kms{\rm km\,s^{-1}}
\def\mach{\mathcal{M}}
\def\angstrom{\mathring{\mathrm{A}}}
\def\grav{\rm G}
\def\gizmo{\textsc{gizmo}}

\section{Introduction}
Within hierarchical models of cosmological structure formation, dwarf galaxies are considered to be the building blocks of larger galaxies. Compared to normal spiral galaxies, dwarfs contain less metals \citep{Tremonti2004,Hunter2012} and have been often employed as laboratories to explore high-redshift conditions. Whether or not this can help disentangling how stars form under early Universe conditions is, however, far to be understood \citep[see][for a different paradigm]{Cowie1996}. 


High redshift objects are mainly traced via the far-infrared (FIR) emission, which represent the best tool to measure star-formation rates (SFRs) in metal-poor environments. Detecting molecular gas in these objects is, in fact, challenging \citep{Leroy2011}. Low metallicities imply a low dust-to-gas ratio, favouring the penetration of photodissociating photons. The main molecular tracer, CO, is then confined in smaller regions, and surrounded by larger photodissociation regions (PDRs), where the emission of strong metal lines (e.g. [CII] and [OI]) is strong, as also proved by observations of [CII] to CO ratios \citep[e.g.][]{Israel1996,Madden1997}. 


In particular, the 158 $\mu$m [CII] transition is the brightest line originating from star-forming galaxies \citep[e.g.][]{Stacey1991,Brauher2008}, and the dominant coolant for neutral atomic gas \citep{Wolfire1995,Wolfire2003}. It has been observed in different types of galaxies, and can also be detected by ALMA in galaxies beyond redshift $z = 6-7$ \citep[see e.g.][]{Capak2015,Maiolino2015}. 

Due to its low ionisation potential (11.26~eV), [CII] traces also other regions of the interstellar medium (ISM), as for instance ionised gas components \citep[see e.g.][]{Goldsmith2012}, it is spatially associated with PDR envelopes surrounding CO, and it correlates with the FIR emission, thus tracing the ultraviolet (UV) field strength.

The role of [CII] as star-formation (SF) tracer, has been widely discussed in the literature \citep{delooze14,herreracamus15}. \citet{delooze2011} reported a nearly linear correlation between SFR and [CII] luminosity with a dispersion of $\sim$0.3~dex for a sample of 24 local, star-forming galaxies, while in a second work \citep{delooze14}
based on the Dwarf Galaxy Survey (DGS hereafter) by \citet{Madden2013}, they concluded that [OI] 63 $\mu$m is a better SF tracer in low-metallicity galaxies. In addition, [CII], normally excited by collision with electrons, atomic, and molecular hydrogen, possesses a lower critical density for collisional excitation ($n_\textrm{crit} \sim 10^3$ cm$^{-3}$ \footnote{This value depends on the collisional partner.}), compared to [OI], which then becomes the dominant coolant above those densities \citep{Kaufman1999}. 

To understand if [CII] can be used as a universal tracer of the star-forming galaxies across time, we should investigate how its emission and chemistry are connected to the metallicity, the far ultra-violet (FUV) background, and its interaction with dust particles. The metallicity, in particular, is a fundamental parameter as it constrains the amount of dust grains, which have a crucial role for the formation of molecules and in shielding the FUV radiation. 

Observational works have also reported the so-called \emph{[CII] line-cooling deficit}, in which the [CII] over the total infrared (TIR) luminosity ratio exhibits large variations. In a recent large survey by \citet{Smith2017} this deficit has been shown within galaxies down to scales of 200~pc. Yet the underlying process (likely local) driving the deficit has not been identified. The impact of harsh radiation in ultra-luminous infrared galaxies (ULIRGs), for instance, might reduce the abundance of small grains and decrease the photoelectric effect (the main heating source), lowering the [CII]/TIR ratio \citep{Luhman2003}. On the other hand, observations of low-metallicity dwarf galaxies \citep{Cormier2015}, have shown opposite behaviour. The lower dust content increases the mean free path of UV photons, causing a UV field dilution over larger spatial scales, which may result in less grain charging, then favouring the photoelectric process.

Together with the [CII] deficit problem, other relevant questions on low-metallicity environments remain open. Determining the role of the different ISM phases on the star formation process, and how these phases are structured are some of those. 

[CII] emission can originate from different regions with contributions from molecular, ionised, and atomic phases of the ISM. For these reasons it might also be employed to study the evolution of the ISM from the warm neutral medium to the cold phase, and could  help determining the physical conditions of the emitting gas. By studying from where [CII] emission originates and comparing it with typical tracers of other ISM phases, we can help finding an answer to the above questions. 
 
\citet{Pineda2013}, via \emph{Herschel} observations of the Galactic plane, have found that dense PDRs are the main sources of the total [CII] emission ($\sim$47 per cent), while only $\sim$ 28 per cent traces CO-dark gas, with most of the emission localised in the spiral arms. \citet{fahrion17}, via SOFIA observations of the dwarf galaxy NGC 4214, reported that roughly 46 per cent of [CII] emission comes from the HI-dominated medium, while only about 9 per cent of the total [CII] emission can be attributed to the cold neutral medium (CNM). In addition, only 25 per cent of the molecular content can be traced by the [CII].

The observations of FIR lines in high-redshift galaxies by ALMA have also been the motivation for the theoretical/numerical community to start mapping the emission lines in galaxies.
However, because of the additional complexity of a proper modelling of non-equilibrium metal chemistry in numerical simulations and the need of very high resolution, most of the studies to date have relied on post-processing techniques, typically based on photoionisation equilibrium models like Cloudy \citep{ferland13}, as in, e.g., \citet{pallottini17b} and \citet{katz19}. Unfortunately, this approach cannot properly describe the dynamical evolution of the species, hence the interplay between chemistry and the other physical processes occurring in the galaxy.

The only cases of a proper investigation of the non-equilibrium state of metal species in simulations to date are the studies by \citet{hu16,hu17}, \citet{hu19} and \citet{richings16}. In \citealt{hu16} papers, they evolved a dwarf galaxy in isolation with an extremely high resolution ($4\, \msun$ per gas particle), employing a simplified network that tracks the primordial species (with only a subset of all the possible ionisation states included) plus only C and O, under the assumption that these metal species can only exist in the molecular form as CO, or separately as C$^+$ and neutral O. On the other hand, \citet{richings16} evolved a sub-L* galaxy in isolation employing a very detailed CO network, with 157 species, but assume a constant metallicity across the entire evolution of the gas. 
Moreover, isolated galaxy simulations lack the ability to self-consistently track the cosmological evolution of the Universe and the role of environment in shaping to the system under scrutiny, like large-scale inflows and mergers, possibly missing important effects.
Here, instead, we build upon our previous study \citet{capelo18} and evolve a dwarf galaxy self-consistently in a cosmological context, with the aim to answer a couple of important questions related to [CII] in galaxies: i) is the $158\,\mu$m [CII] line a valid star formation rate tracer in local galaxies and across time? ii) To which extent [CII] traces different components of the ISM?

\section{Numerical setup}
\label{sec:setup}
In this study, we consider the evolution of a dwarf galaxy with a halo mass of $\sim 4\times 10^{10}~\msun$ at $z=0$ by means of a zoom-in cosmological simulation performed with the hydrodynamic code \gizmo{} \citep{hopkins15}, descendant from \textsc{gadget3} and \textsc{gadget2} \citep{springel05}. For hydrodynamics, we employ the fully-Lagrangian mesh-less finite mass (MFM) method.

\subsection{Sub-grid physical processes included}
To model the main physical processes responsible for galaxy formation and evolution, we employ the following sub-grid prescriptions.

\begin{itemize}
\item The metal chemical network and cooling/heating processes from \citet{capelo18}, modelled via \textsc{krome} \citep{grassi14},
that included non-equilibrium chemistry for hydrogen and helium species, and the six most relevant metal species responsible for the cooling in the ISM \citep{richings14,bovino16} (C,C$^+$,O,O$^+$,Si,Si$^+$,Si$^{++}$). 
Compared to \citet{capelo18}, we improved our chemical network by including three-body reactions involving H$_2$ \citep{glover08,forrey13}, H$^+_2$ collisional dissociation by H, and H$^-$ collisional detachment by He \citep{glover09} for completeness. 
As in \citet{lupi18} and \citet{lupi19a}, we also include a clumping factor $C_\rho = 1+b^2\mach^2$ in the H$_2$ formation rate on dust, where $\mach$ is the Mach number and $b=0.4$ is a parameter accounting for the ratio between solenoidal and compressive modes \citep{federrath12}.
\item A stochastic SF prescription that converts gas particles into stellar particles, where the SFR density is defined as
\begin{equation}
\dot{\rho}_{\rm SF} = \varepsilon \frac{\rho_{\rm g}}{t_{\rm ff}},
\end{equation}
with $\rho_{\rm g}$ the local gas density, $t_{\rm ff}=\sqrt{3{\rm \pi}/(32 G \rho_{\rm g})}$ the free-fall time, $G$ the gravitational constant, and $\varepsilon$ the SF efficiency parameter, defined as \citep{padoan12}
\begin{equation}
\varepsilon=\varepsilon_\star\exp\left(-1.6\frac{t_{\rm ff}}{t_{\rm dyn}}\right)
\label{eq:sfeff}
\end{equation}
where $\varepsilon_\star=0.9$ is the local SF efficiency \citep{semenov16}, $t_{\rm dyn}=L/(2\sigma_{\rm eff})$ is the dynamical time of the star-forming cloud, and $L$ is the cloud size. \footnote{As in \citet{lupi18} and \citet{lupi19a}, $L\approx 0.5 h$ is the particle grid-equivalent size, with $h$ the particle smoothing length, $\sigma_{\rm eff}\equiv L\sqrt{\|\nabla\otimes\mathbf{v}\|^2+(c_{\rm s}/L)^2}$ is the total support against gravitational collapse, with $\nabla\otimes\mathbf{v}$ the velocity gradient in the cloud and $c_{\rm s}$ the sound speed.}
As was the case in \citet{lupi18}, we employ a very low SF density threshold $\rho_{\rm g}/ m\rm _H = 1\, cm^{-3}$ to avoid wasting time computing the SF rate in low-density, unbound regions that are never expected to form stars.
 
\item Stellar radiation is implemented as in model {\it b}  of \citet{lupi18}, that showed the best agreement with on-the-fly radiative transfer calculations. 
With respect to \citet{lupi18}, here we update the shielding calculation around the gas cell with the best local approximation by \citet{safranekshrader17}, where the gas temperature used to estimate the Jeans length is capped at 40~K.
In particular, unlike several other studies, photo-electric heating is self-consistently computed from the total flux reaching the gas cell, as in \citet{hu17}.

\item Supernova feedback is based on the mechanical feedback prescription in \citet{lupi19a}, a variation of the publicly available implementation by \citet{hopkins18b}.
As described in \citet{geen15}, the preprocessing by stellar radiation close to the stellar sources is able to reduce the gas density before SN events. This leads to an increase of the terminal momentum for high gas densities,
corresponding to $p_{\rm fin,max} = 4.2\times 10^5 E_{\rm SN, 51}^{16/17} \zt^{-0.14}\rm\, \msun~km~s^{-1}$, where $E_{51}$ is the SN energy $E_{\rm SN}$ in units of $10^{51}$~ergs and $\zt=\max\{Z/Z_\odot,0.01\}$ with $Z_\odot$ the solar metallicity. Here, since we cannot properly model the radiative effect within molecular clouds, we include this additional boost by rescaling the injected momentum by a factor $f_{\rm boost} = \max\{p_{\rm fin,max}/p_{\rm fin},1\}$, where $p_{\rm fin} = \sqrt{2E_{\rm SN} M_{\rm cool}}$ is the `standard' terminal momentum, with $M_{\rm cool}$ the cooling mass from \citet{martizzi15}. 

Motivated by \citet{semenov18} and \citet{lupi19a}, we also include an additional fiducial boost to the radial momentum of a factor of two.

In our model, stellar particles represent an entire stellar population following a Kroupa initial mass function \citep[IMF;][]{kroupa01}, that release mass and energy via stellar winds and type II/Ia SN events. The type II SN and stellar wind rates have been computed via \textsc{starburst99} \citep{leitherer99}, assuming that only stars with masses between 8~and~40~$\msun$ can explode as type II SNe, whereas those in the range $40-100\,\msun$ directly collapse into a black hole. 
As in \citet{lupi19a}, SN explosions are modelled as discrete events releasing $10^{51}$~erg of energy and the IMF-averaged ejecta masses, respectively a total ejecta mass $M_{\rm ej}=10.61\zt^{-0.096}~\msun$, an oxygen mass $M_{\rm oxy}=1.021\,\msun$ and an iron mass $M_{\rm iron}=0.106\,\msun$. The total metal mass injected per single event, assuming solar ratio for the Oxygen and Iron groups, respectively, corresponds to $M_{\rm Z} = 2.09M_{\rm oxy}+1.06M_{\rm iron}=2.246\,\msun$.  For type Ia SNe, instead, we employ the delay-time distribution from 
\citet{maoz12}, spanning the stellar age interval 0.1-10~Gyr, and we inject $10^{51}$~erg, $M_{\rm ej}=1.4\,\msun$, $M_{\rm oxy}=0.14\,\msun$, and $M_{\rm iron}=0.63\,\msun$, with $M_{\rm Z}=0.9604\,\msun$. 
The \textsc{starburst99} rates for type II SNe can be well modelled by a single power-law defined as
\begin{equation}
\frac{\dot{N}_{\rm SN}}{\rm Myr^{-1}~\msun^{-1}} =6.8\times  10^{-3}(t_{\rm Myr}/t_{\rm max,Myr})^{-0.648}/t_{\rm max,Myr},
\end{equation}
for $t_{\rm min,Myr}<t_{\rm Myr}<t_{\rm max,Myr}$, where $t_{\rm Myr}$ is the stellar population age in Myr, $t_{\rm min,Myr}=5.09$ is the typical lifetime of an $40~\msun$ star and $t_{\rm max,Myr}=38.1$ is the typical lifetime of a $8~\msun$ star.
For type Ia SNe, the rate can be defined as
\begin{equation}
\frac{\dot{N}_{\rm SNIa}}{\rm Myr^{-1}~\msun^{-1}} = 2.82\times 10^{-4} t_{\rm Myr}^{-1}
\end{equation}
\item Winds are implemented by distributing the wind mass and the wind energy in thermal form only, assuming a continuous damping taken from the \textsc{starburst99} rates, that can be fitted as 
\begin{equation}
\frac{\dot{M}_{\rm w}}{\rm Myr^{-1}} =
\begin{cases}
1.9\times10^{-3}+6.5\times10^{-4}t_{\rm Myr} & 0.0\leq t_{\rm Myr}<2.7\\
2.58\times10^{-2} & 2.7\leq t_{\rm Myr}<3.3\\
2.58\times10^{-2}m_x^{(\log t_{\rm Myr}-0.52)/1.48} & 3.3\leq t_{\rm Myr} <t_x\\
4\times10^{-3}m_x^{\log(t_{\rm Myr}/t_x)/\log(25/t_x)}+\\ 2.58\times10^{-3}m_x^{(\log t_{\rm Myr}-0.52)/1.48} & t_x<t_{\rm Myr}<100\\
5\times 10^{-4}(t_{\rm Myr}/100)^{-1.24} & t_{\rm Myr}>100\\
\end{cases}
\end{equation} 
with $m_x=1.94\times10^{-3}$ and $t_x=8.7\min\{1,(0.5+\zt/2)^{0.4}\}$ for the mass and 
\begin{equation}
\frac{\dot{E}_{\rm w}}{\rm L_\odot~\msun^{-1}}=
\begin{cases}
2.32 & 0.0\leq t_{\rm Myr}<2.7\\
4.83 & 2.7\leq t_{\rm Myr} < 3.3\\
4.83f_x^{(\log t_{\rm Myr}-0.54)/0.6} & 3.3\leq t_{\rm Myr} <100\\
4.1\times 10^{-5} (t_{\rm Myr}/100)^{-1.24}  & t_{\rm Myr}>100\\
\end{cases}
\end{equation}
where $f_x=1.23\times 10^{-3}$. We note that the wind power for old stars ($t_{\rm Myr}>100$, producing slow winds) is simply computed as $\dot{E} = 0.5 \dot{M} v_{\rm eff}^2$, with $v_{\rm eff}=30\rm~km~s^{-1}$.
While the metallicity dependence for old stars is very weak, for young stars we rescale the rates according to the stellar population metallicity as $\dot{M}_{{\rm w},Z} = \dot{M}_{\rm w} \zt^{0.8}$ and $\dot{E}_{{\rm w},Z}=\dot{E}_{\rm w}\zt^{1.06}$ if $t_{\rm Myr}<2.7$, and as $\dot{M}_{{\rm w},Z} = \dot{M}_{\rm w} \min\{1,\zt^{0.8}\}$ and $\dot{E}_{{\rm w},Z}=\dot{E}_{\rm w}\min\{1,\zt^{1.06}\}$ if $2.7\leq t_{\rm Myr}<t_{\rm max,Myr}$.

\end{itemize}

During a Hubble time, the ensemble of stellar feedback prescriptions results in a mass recycling of about 40 per cent \citep[see][and references therein]{kim14}. 
In order to resolve single SN events in our simulations, we also add a time-step limiter for stellar particles (see \citealt{lupi19a} for details).

\subsection{Initial conditions}
The initial conditions consist of two high-resolution regions, around halos with $M_{\rm halo}\sim 4\times 10^{10}~\msun$ and $M_{\rm halo}\sim 10^{11}~\msun$ at $z=0$, H1 and H2 hereafter, in a box of $10\rm\, Mpc\,h^{-1}$~comoving. We generated the initial conditions with \textsc{MUSIC} \citep{hahn13}, assuming the {most recent} cosmological parameter estimates from PLANCK \citep{planck18}, where $\Omega_{\rm m}=0.3119$, $\Omega_\Lambda=0.6881$, $\Omega_{\rm b}=0.04669$, and $H_0=67.66\rm~km s^{-1} Mpc^{-1}$.
We first ran a coarse level dark-matter only simulation (256$^3$) and selected the two haloes in the desired mass range, by enforcing a strong isolation criterion, i.e. no haloes more massive than half the target mass were present within a sphere of $3R_{\rm vir}$, with $R_{\rm vir}$ the halo virial radius. 

To identify the haloes in the simulation we employed \textsc{amiga halo finder} \citep[][AHF hereafter]{knollmann09}. To refine the target region, we followed the procedure described by \citet{fiacconi17} and \citet{lupi19b} by recursively selecting and tracing back in time all the particles within $3R_{\rm vir}$ at $z=0$.

The mass resolution in the high-resolution region is $\sim1.2\times 10^4~\msun$ for dark matter particles and $\sim2.2\times 10^3~\msun$ for gas and star particles.
The gravitational softening for dark matter and stars in the high-resolution region is fixed at 60 and 7~pc~h$^{-1}$, respectively, whereas we employ fully adaptive gravitational softening for the gas, which ensures both hydrodynamics and gravity are computed assuming the same mass distribution within the kernel. The minimum allowed softening, which sets the maximum resolution of the simulation, is set to 0.5~pc~h$^{-1}$, corresponding to an inter-particle spacing of $\sim 1$~pc.

\section{Results}
\label{sec:results}
Here, we present our results, focusing first on the galaxy properties at the final redshift of the runs, i.e. $z=0$ for H1 and $z=0.4$ for H2, and later on the redshift evolution. For all the analysis, we only consider particles within a sphere of 0.2$R_{\rm vir}$, to exclude any possible satellites around the target galaxies. The main properties are reported in Table\ref{tab:results}.

\begin{table}
    \centering
    \caption{Main properties of our simulated galaxies at their final redshift. The first column is the halo mass, the second and the third are the stellar mass and the SFR, the fourth is the half-mass radius obtained from the 2D maps (first value) or from the 3D distribution (second value), the fifth the stellar metallicity, and the sixth the total [CII] luminosity.}
\begin{tabular}{cccccc}
        \hline\hline
        $M_{\rm halo}$ & $M_{\rm star}$ & SFR & $R_{50}$ & $Z_{\rm star}$ & $L_{\rm CII}$ \\
        ($10^9\msun$) & ($10^9\msun$) & ($\msun\, \rm yr^{-1}$) & ($\rm kpc$) &($Z_\odot$) & ($10^5 L_\odot$)\\
        \hline
         $38.2$ & $0.24$ & $4.4\times 10^{-3}$ & 1.2/1.4 & 0.27 & $0.20$\\
         $70.8$ & $1.08$ & $8.2\times 10^{-2}$ & 1.5/1.7 & 0.41 & $4.94$\\
     \hline\hline
    
    \end{tabular}
    \label{tab:results}
\end{table}

\begin{figure}
\centering
\includegraphics[width=\columnwidth]{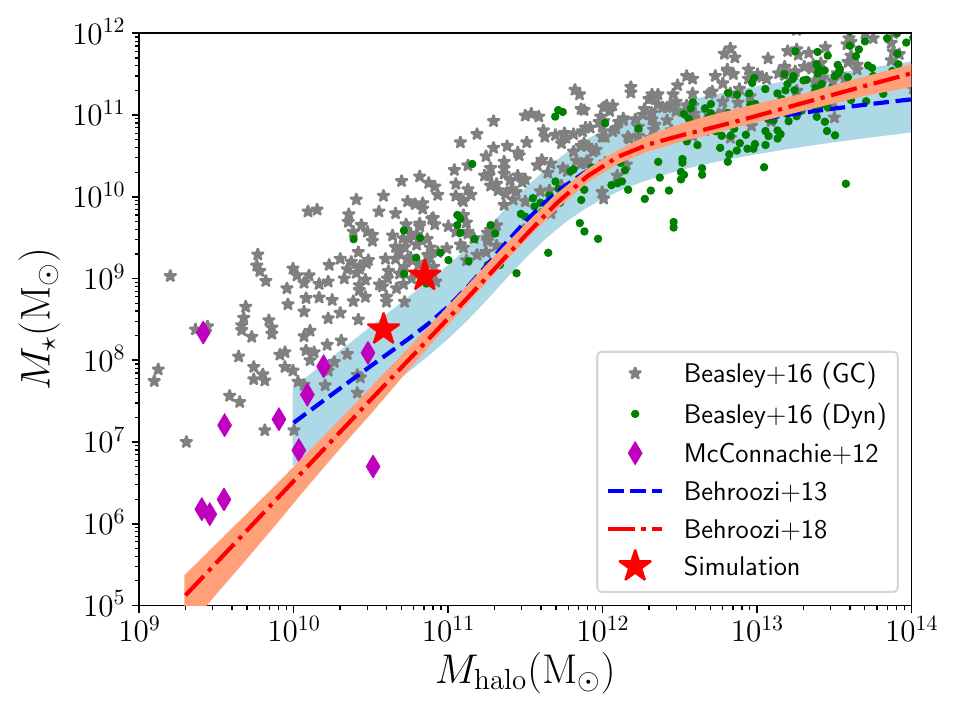}
\caption{Stellar-to-halo mass relation in the simulations, compared with observations and semi-empirical models. The grey stars correspond to the halo mass from globular cluster measurement from \citet{beasley16}, the green dots to their dynamical mass estimates, and the magenta diamonds to the dwarf galaxies in the Local Group by \citet{mcconnachie12}. The blue dashed line is the model by \citet{behroozi13} and the red one is the updated model from \citet{behroozi18}, both at $z=0$, with the light-blue and orange shaded areas representing the 3$\sigma$ uncertainty in the models. Our simulations, shown as  red stars, are in good agreement with the data, well within the uncertainties.}
\label{fig:smhm}
\end{figure}

In Fig.~\ref{fig:smhm}, we show the stellar-to-halo mass relation for our dwarf galaxies (red stars), compared with observational data from \citet{beasley16}, where the halo mass is estimated from globular cluster counting (grey stars) and dynamical measure (green dots), Local Group dwarf galaxies from \citet[][; magenta diamonds]{mcconnachie12}, and the semi-empirical models by \citet{behroozi13} and \citet{behroozi18}, reported as a blue dashed line and a red dot-dashed line, respectively. The shaded areas correspond to 3$\sigma$ uncertainty in the models. Our  galaxies are in good agreement with both observations and theoretical models, lying within the scatter of the data. 

\begin{figure}
\centering
\includegraphics[width=\columnwidth]{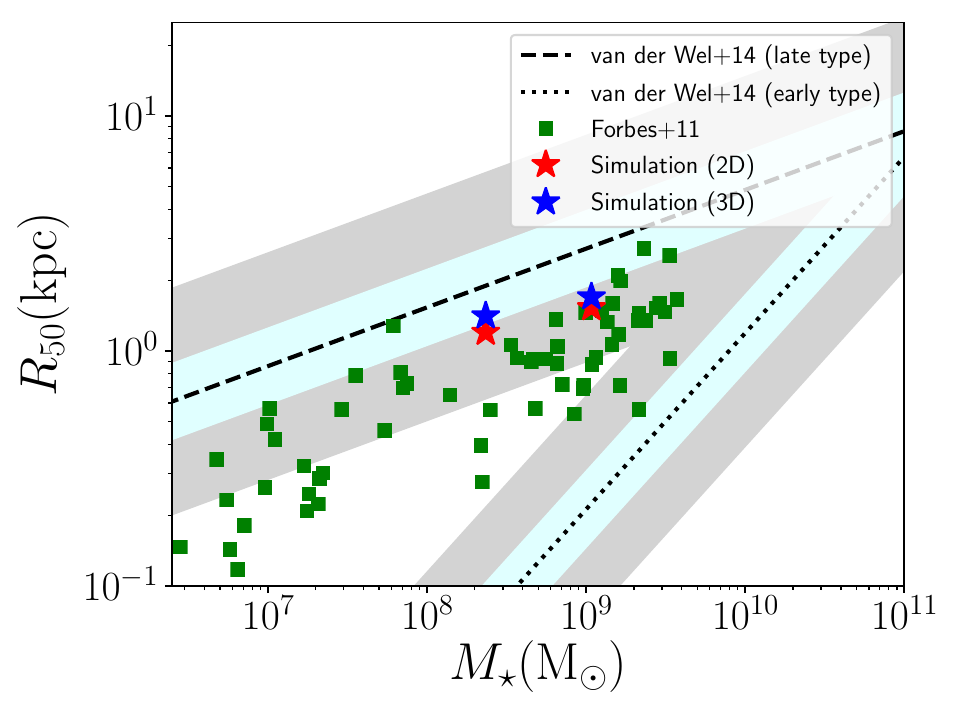}
\caption{Stellar mass-size relation in the simulations, compared with observations from the 3D+HST+CANDELS survey for $0.25<z<0.75$ \citep[black dashed lines and corresponding shaded areas tracing 1- and $3\sigma$ uncertainties][]{vanderwel14} and the dwarf spheroidal galaxies from \citet{forbes11}, shown as green squares. Our simulations agree very well with the data, almost exactly following the trend observed for dwarf spheroidal galaxies.}
\label{fig:msize}
\end{figure}

\begin{figure}
\centering
\includegraphics[width=\columnwidth]{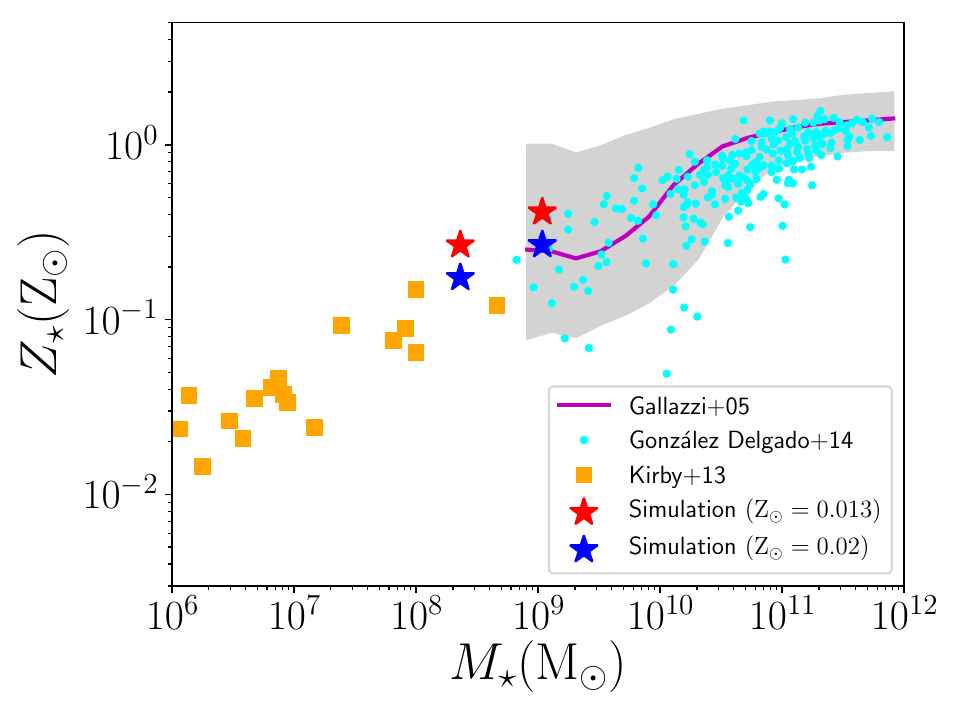}
\caption{Stellar mass-metallicity relation in the simulations, compared with observations of local dwarfs \citep[][orange squares]{kirby13}, normal galaxies of the CALIFA survey \citep[][cyan dots]{delgado14}, and those by \citet{gallazzi05}, for which we show the median profile as a magenta solid line and the uncertainty within the 16th and 84th percentile as a grey shaded area. Our simulation results are shown as stars, assuming different definitions for the solar metallicity, i.e. red for \citet{asplund09} and blue for \citet{anders89}. Our results are in very good agreement with local dwarfs, and also consistent with the low-mass end of the CALIFA survey.}
\label{fig:massz}
\end{figure}

In Fig.~\ref{fig:msize}, we show the half-mass radius of the two galaxies as a function of the stellar mass, compared with the dwarf spheroidal galaxies from \citet{forbes11}, shown as green squares, and the results from the 3D+HST+CANDELS survey \citep{vanderwel14}, reported through the best-fit to late type (black dashed line) and early type (black dotted line) galaxies at $0.25<z<0.75$. The cyan and grey dashed areas correspond to 1$\sigma$ and 3$\sigma$ uncertainty, respectively. Our simulation{\rm s are} reported as blue stars (computing the size in 3D radial bins) and red stars (computing the size from the 2D face-on projection), and exhibit a very good agreement with the observational data, in particular with the dwarf spheroidals.
In Fig.~\ref{fig:massz}, we report the stellar mass--metallicity relation, compared with the local dwarfs by \citet[][orange squares]{kirby13}, the CALIFA survey from \citet[][cyan dots]{delgado14}, and the median from \citet{gallazzi05}, shown as a magenta line, with the grey shaded area corresponding to the 16th and 84th percentiles of the distribution. Because of the different assumptions on the solar metallicities when deriving these relations, that are $\rm Z_\odot=0.02$ \citet{anders89} or $\rm Z_\odot=0.013$ \citep{asplund09}, we show the expected metallicity by employing both definitions, respectively as blue and red stars. Independent of the definition for $\rm Z_\odot$, our results are in very good agreement with local dwarfs, and are very close to the median by \citet{gallazzi05}, despite the stellar mass in the H1 run is just outside the available data interval.

\subsection{SF history and the Schmidt--Kennicutt relation}
\begin{figure}
\centering
\includegraphics[width=\columnwidth]{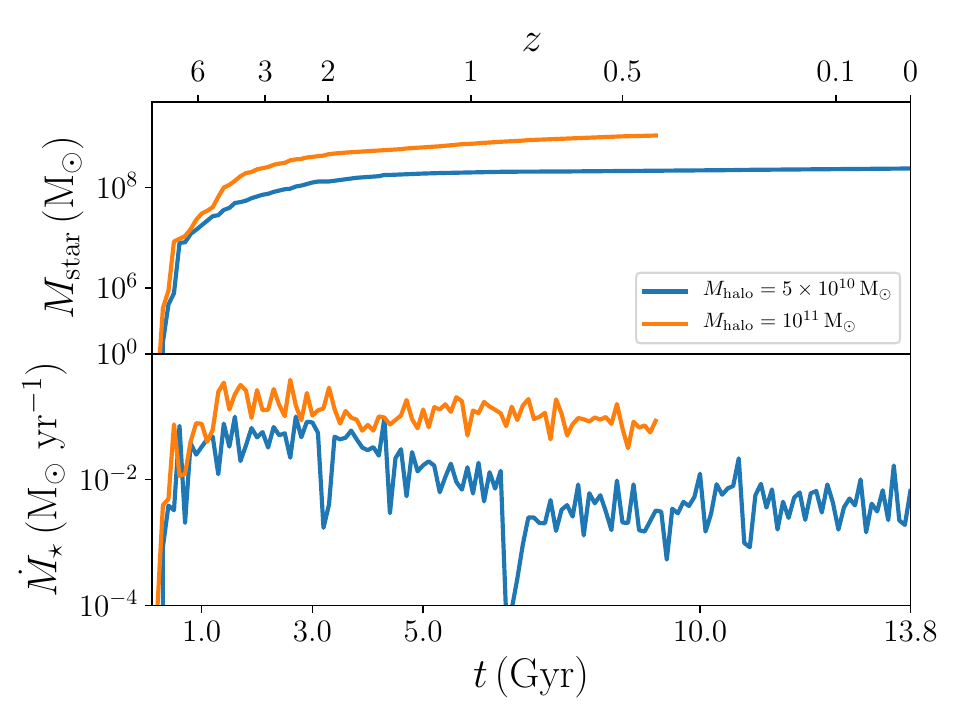}
\caption{SF history of the target galaxies in bins of 100~Myr. The top panel shows the stellar mass as a function of time, whereas the bottom panel shows the SFR. In H1, most of the stellar mass is  built up in the first 3~Gyr (above $z=2$), when the average SFR exceeds 0.01~$\msun~\rm yr^{-1}$, whereas H2 assembles less than half its mass by this time. At later times, instead, the lower gas densities, combined with the stronger radiative heating (UV background and local source) result in a suppression of the SFR at about $\dot{M}_{\rm star}\sim 10^{-3}~\msun~\rm yr^{-1}$ for H1, while H2 continues to form stars at a steady pace of about $0.1~\msun~\rm yr^{-1}$.}
\label{fig:sfh}
\end{figure}

Here, we discuss the SF history of the galaxies and we compare our results with the spatially-resolved local correlation between the gas and the SFR surface densities, the so-called Schmidt--Kennicutt relation \citep[KS hereafter;][]{schmidt59,kennicutt98}.
In Fig.~\ref{fig:sfh}, we report the SF history of the target galaxies as a function of time/redshift. In H1, stars begin to form already at $z>6$, with a SFR of about 0.03~$\msun\rm~yr^{-1}$ that lasts until $z\sim 2$. These `high' SFRs are responsible for most of the SF occurring in the galaxy, that almost reaches its final mass by $z=2$. At later times, instead, the typically lower gas densities, together with the radiative heating from the uniform UV background and the local sources in the galaxy result in the suppression of the SFR at about $\dot{M}_{\rm star}\sim 10^{-3}~\msun~\rm yr^{-1}$. In H2, instead, the SFR does not vary significantly with time, only slightly declining from the peak of $\sim 0.3~\msun\rm~yr^{-1}$ at $z\sim 5$ to about $0.1~\msun\rm~yr^{-1}$. This explains the steady increase in stellar mass, that reaches $10^9\rm\msun$ by $z=0.4$. In this second case, less than half of the mass is assembled before $z=2$.

\begin{figure*}
\centering
\includegraphics[width=\columnwidth]{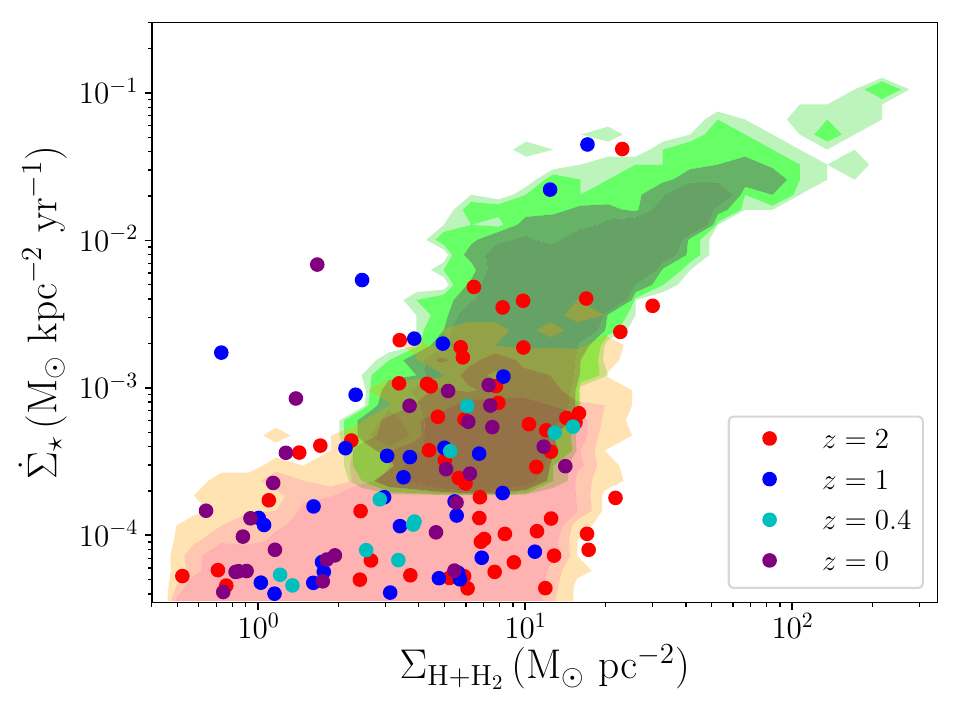}
\includegraphics[width=\columnwidth]{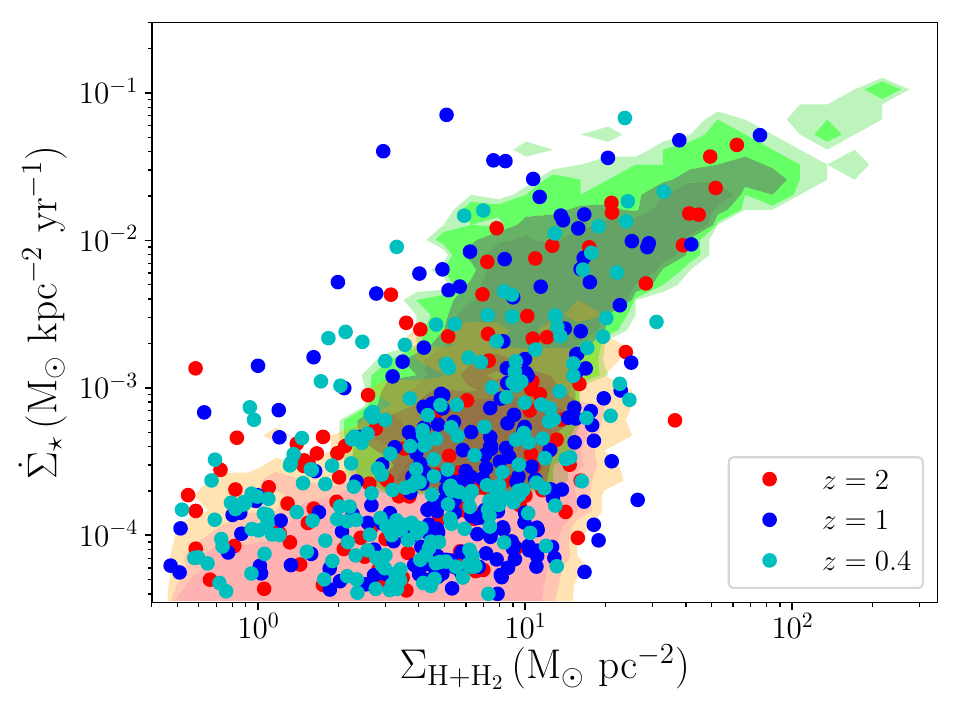}
\caption{KS relation in total gas at $z=0$ for the target galaxies (H1 on the left and H2 on the right) in our simulations (filled dots) at $z=2,1,0.4$,~and~0 (for H1 only), compared with the observations by \citet{bigiel08} (green contours) and \citet{bigiel10} (red contours). The increased momentum injected by SNe is able to heat the gas up more effectively, resulting in a good agreement between simulations and observations.}
\label{fig:ks}
\end{figure*}

Now, in Fig.~\ref{fig:ks} we show the KS relation in total gas (H+H$_2$), computed in concentric annuli 500~pc wide, for $z=2$ (red dots), $z=1$ (blue dots), $z=0.5$ (green dots), and $z=0$ (cyan dots). 
The SFR is estimated from the far UV luminosity $L_{\rm FUV}$ of the stellar particles in the simulation, assuming the calibration by \citet{salim07}, i.e. $\dot{M}_{\rm star} = 0.6\times 10^{-28} L_{\rm FUV}$.\footnote{We also computed the SFR directly from stellar particles in the simulations, finding consistent results.} The filled contours correspond to the observations by \citet{bigiel08} (green color-scale) and \citet{bigiel10} (red color-scale), where the different colours correspond to 2,5, and 10 points per bin. Compared to \citet{lupi19a}, here the agreement between the simulation and observations is very good.

In dwarf galaxies, SF mostly occurs in the HI-dominated regime ($\Sigma_{\rm H+H_2}\lesssim 10\rm\,\msun pc^{-2}$, where H$_2$ is strongly subdominant \citep[e.g.][]{hu17}, hence the molecular KS correlation is not expected to always hold, especially for very-small galaxies. Indeed, for H1, most of our data lies below the transition surface density, whereas for H2 more data are found above this threshold.

\begin{figure*}
\centering
\includegraphics[width=\columnwidth]{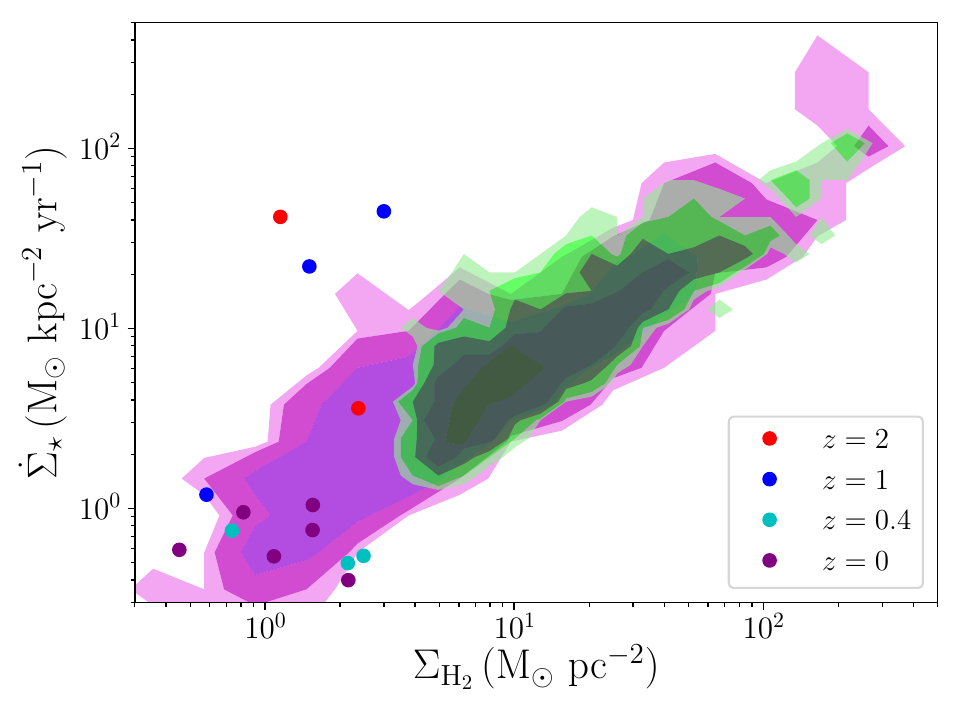}
\includegraphics[width=\columnwidth]{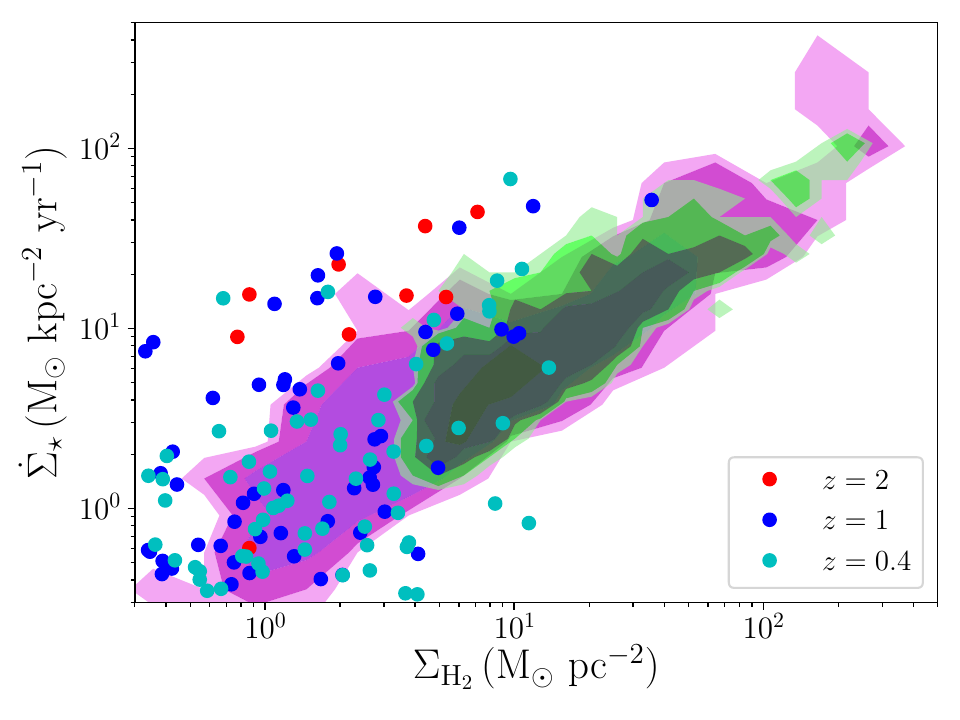}
\caption{KS relation in molecular gas at $z=0$ for the target galaxies in our simulations (filled dots) ad $z=2,1,0.4$,~and~0 (for H1 only), compared with the observations by \citet{bigiel08} (green contours) and \citet{schruba11} (magenta contours). Despite the few points, the simulation is in reasonable agreement with the observations, except for three points at higher redshift that lie above the correlation (due to SNe quickly evacuating the star-forming high-density gas).}
\label{fig:ksh2}
\end{figure*}

We report the molecular KS relation in Fig.~\ref{fig:ksh2}, comparing our simulations with observations by \citet{bigiel08} (green contours) and \citet{schruba11} (magenta contours).
In H1, we observe that only a few regions exhibit an agreement with this correlation, whereas the others show a deficit in H$_2$, as expected. This deficit is due to the strong impact of SNe onto the dense gas, that is quickly evacuated, on time-scales much shorter than those traced by the far UV emission.
In H2, instead, where SNe start to become less effective, most of the examined patches lie within the observational scatter, with only a few of them lying above (due to the still strong impact of SNe) or even below (where the formed H$_2$ has not been converted yet into stars, likely because of the moderately lower densities and stronger turbulent support).
In general, our simulations show a good agreement with observations, especially at low redshift ($z<=0.4$), when most of the patches lie at the centre of the observational contours.  

\subsection{The [CII]--SFR correlation}
\begin{figure*}
    \centering
    \includegraphics[width=\textwidth,trim=3.5cm 3cm 2.5cm 0cm,clip]{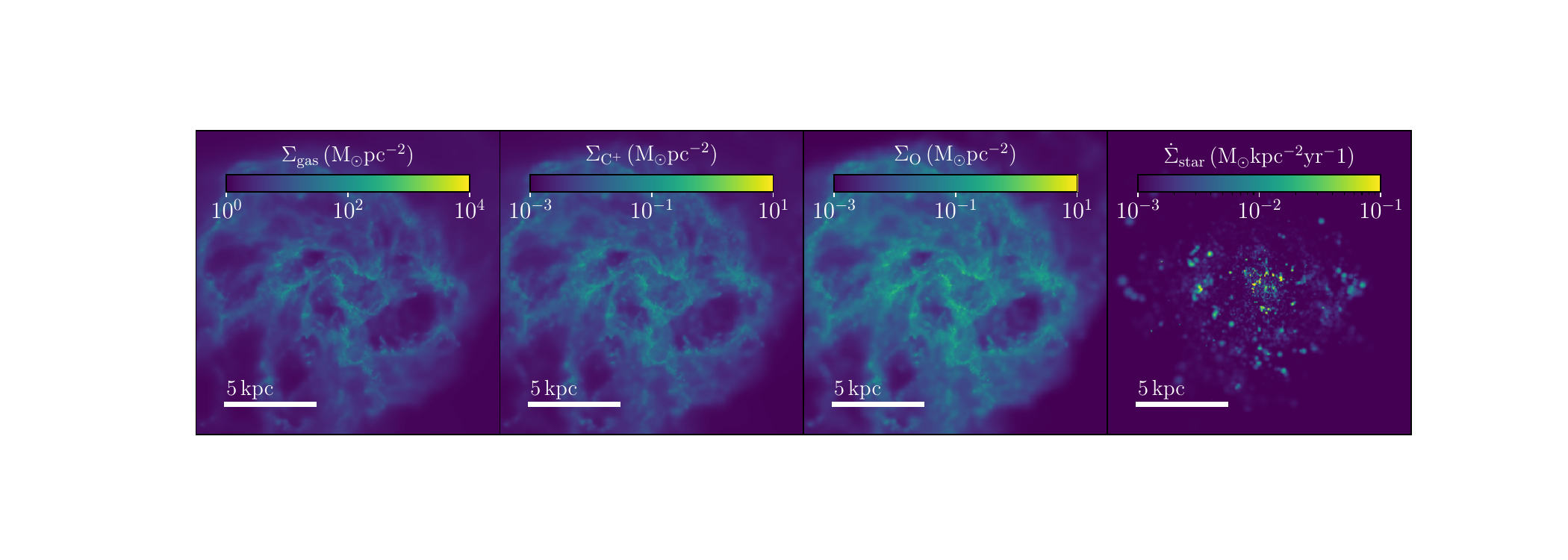}
    \caption{Surface density maps of total gas (leftmost panel), C$^+$ (middle-left panel), O (middle-right panel), and SFR (rightmost panel) for our H2 run at $z=0.4$. Although strongly perturbed by SN explosions, the galaxy clearly exhibits a disk-like distribution, with C$^+$ and especially O showing a stronger contrast relative to the total gas in dense regions, i.e. the regions that also produce the bulk of the emissivity of the galaxy. Consistently, in the rightmost panel SF is observed mainly in these dense regions, that likely correspond to molecular clouds.}
    \label{fig:map}
\end{figure*}

\begin{figure}
\centering
\includegraphics[width=\columnwidth]{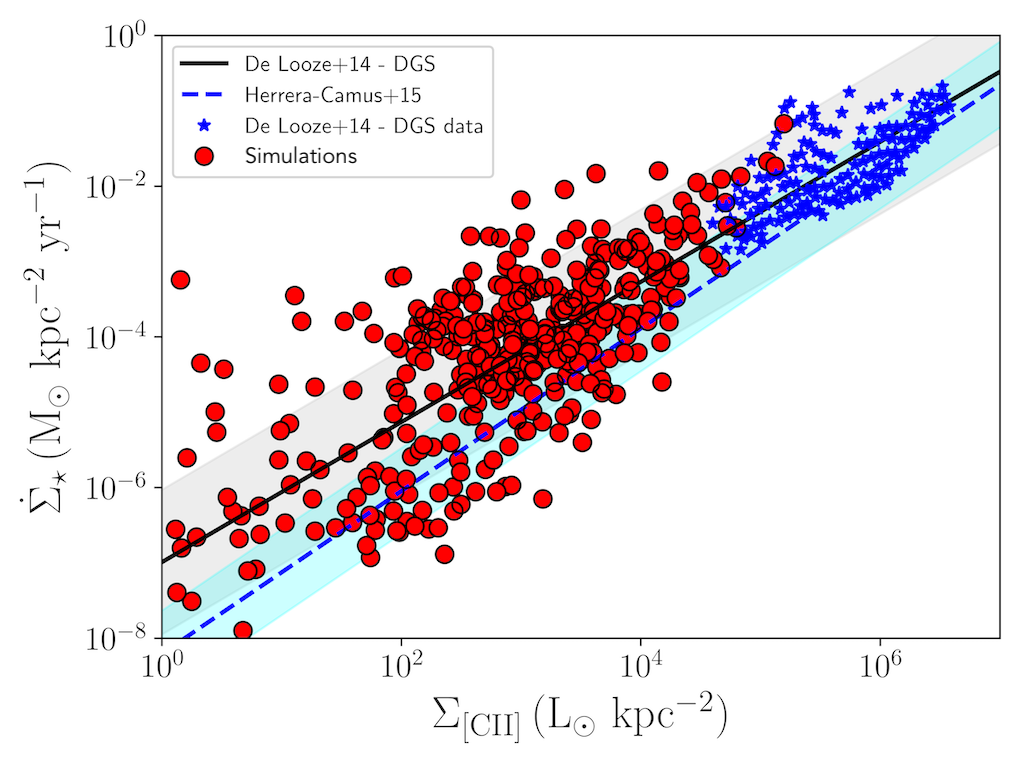}
\caption{Spatially-resolved [CII]--SFR correlation in the simulations (red dots), compared with the best-fits for the DGS galaxies by \citet{delooze14}, shown as a black solid line, and for the KINGFISH sample by \citet{herreracamus15}, shown as a blue dashed line, where the grey and cyan shaded areas corresponds to the $3\sigma$ uncertainties. The DGS data is shown as purple stars. Our simulation data, binned in square patches 500~pc wide, exhibits a good agreement with both correlations by \citet{delooze14} and \citet{herreracamus15}, despite having a large scatter.}
\label{fig:ciisfr_res}
\end{figure}

After having shown that our galaxy agrees well with observational constraints, we now focus on the [CII] emission from the galaxy, in the aim at assessing whether we can reproduce the expected correlation by \citet{delooze14}, and its evolution with metallicity.
Thanks to the chemical network we have in our simulations, we can directly follow the abundances of different elements during the cosmic evolution. An example of the chemical species surface density is shown in Fig.~\ref{fig:map} for our H2 galaxy at $z=0.4$. The leftmost panel corresponds to the total gas surface density, the middle-left one to the the C$^+$ one,  the middle-right one to the O one, and the rightmost to the SFR surface density derived from the FUV emission, respectively.
We can clearly see that on one hand O appears more abundant in dense and cold regions compared to C$^+$, that instead more effectively recombines to neutral C at high densities, suppressing the expected [CII] emission. On the other hand, O can remain in its neutral form also at intermediate densities, hence partially contributing to the total OI emission, where we expect most of the [CII] emission being produced. Consistently with this distribution, star-forming gas in the galaxy appears concentrated in the densest regions, likely corresponding to molecular clouds.
By comparing the relative abundance of C$^+$ and O, the latter appears to be a better tracer of SF compared to the former. However, a proper comparison cannot be made on the basis of abundances alone, but requires the estimation of the effective FIR lines emissivity, that is strongly dependent on the thermodynamic state of the gas and the relative abundance of colliders able to excite the line transition \citep[see][for details]{grassi14}. By using \textsc{krome}, we can estimate the emission of the [CII] and the [OI]$_{63}$ lines from our chemical network, and directly compare them with observations. Here we focus on the [CII] only, and defer the reader to Appendix~\ref{app:oirelation} for a similar analysis applied to the [OI]$_{63}$ case.
In Fig.~\ref{fig:ciisfr_res}, we compare our simulations (red dots) with the spatially-resolved correlation by \citet{delooze14} for dwarf galaxies, shown as a black solid line (best-fit) and as blue dots (data), and the best-fit for normal, nearby galaxies from the KINGFISH sample by \citet{herreracamus15}, shown as a blue dashed line (with the $3\sigma$ uncertainties identified by the grey and cyan shaded areas, respectively). The simulation data is binned in square patches 500~pc wide, with the SFR derived from the far UV luminosity and the [CII] luminosity from the corresponding cooling rate estimated by \textsc{krome} at 158~$\mu$m. Our simulation shows a good agreement with the correlations by \citet{delooze14} and \citet{herreracamus15}. The combined scatter of the two runs is slightly larger than that observed, but observations in this luminosity range are still missing, and future observations with higher sensitivity will be crucial to improve the fit and validate/confute our results.

\begin{figure}
\centering
\includegraphics[width=\columnwidth]{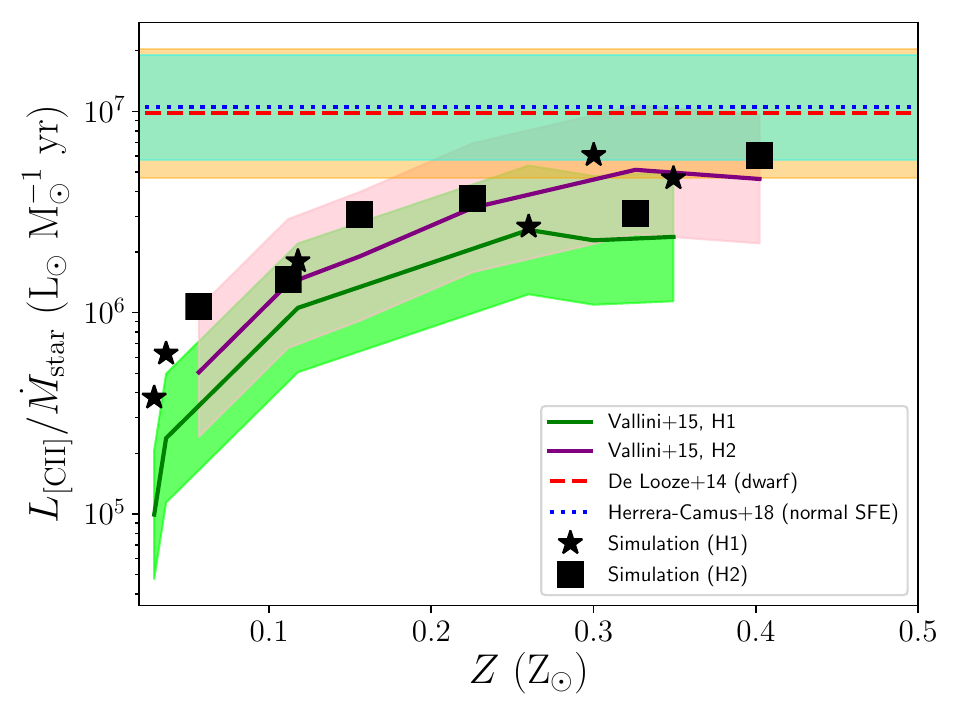}
\caption{Redshift evolution of the ratio $L_{\rm [CII]}/\dot{M}_{\rm star}$ with redshift, compared with the model predictions by \citet[][green solid line]{vallini15}, the observed correlation by \citet{delooze14} for the DGS sample (red dashed line), and the correlation by \citet{herreracamus18} for galaxies with normal SFE (blue dotted line). The shaded areas correspond to the $1\sigma$ uncertainties, i.e. 0.32~and~0.26~dex, respectively. Our simulations are reported as black stars (H1) and squares (H2), as a function of the average gas metallicity in the galaxy during its redshift evolution. The results suggest a clear metallicity evolution for the correlation, in agreement with \citet{vallini15}.}
\label{fig:ciisfr_evo}
\end{figure}

While a spatially-resolved correlation can be obtained for local galaxies with high-resolution observations, at high redshift this is not achievable anymore, and the integrated correlation is employed instead. Now, we investigate the metallicity evolution of the integrated correlation, by computing the ratio between the total [CII] luminosity $L_{\rm [CII]}$ and the SFR $\dot{M}_{\rm star}$. Since we follow a single galaxy only, the metallicity evolution directly follows from the redshift evolution, and we cannot disentangle any distinct effect among the two. According to \citet{vallini15}, at $z\gsim 4$, the higher CMB temperature goes into equilibrium with the spin temperature of the transition powering the [CII] line, and the stronger UV pumping changes the state populations. This results in a  suppression of the [CII] luminosity. At lower redshift, this effect disappears, and the [CII]-SFR correlations loses any intrinsic dependence on redshift. For this reason, we can safely assume here that the only evolution of the correlation expected in our simulations is due to the metallicity evolution of the galaxy. In Fig.~\ref{fig:ciisfr_evo}, we show the ratio at different metallicities for the two galaxies, 
compared with the predictions by \citet{vallini15} and the best-fits from \citet{delooze14} for the DGS galaxies (red dashed line) and from \citet{herreracamus18} for galaxies with normal SF efficiency (SFE; blue dotted line). The shaded areas correspond to the scatter in the relations, i.e. 0.32~and~0.26~dex, respectively. The two runs, shown as black stars (H1) and squares (H2),
 exhibit a clear evolution with metallicity, with a surprisingly good agreement with the theoretical model by \citet{vallini15}, and only approaching the \citet{delooze14} and \citet{herreracamus18} data at $z=0$ and $Z\sim 0.3-0.5\rm~Z_\odot$.

Unfortunately, due to the scatter in the observed [CII]--SFR correlation, the evolutionary trend we observed in the simulation can easily remain hidden within the uncertainty. Indeed, if we consider the $3\sigma$ scatter in the sample, with $\sigma\approx 0.3$, all our data points fall within the uncertainty range. 

\subsection{The [CII] emitting gas distribution}
\label{sec:ciiphase}

\begin{figure}
\centering
\includegraphics[width=\columnwidth]{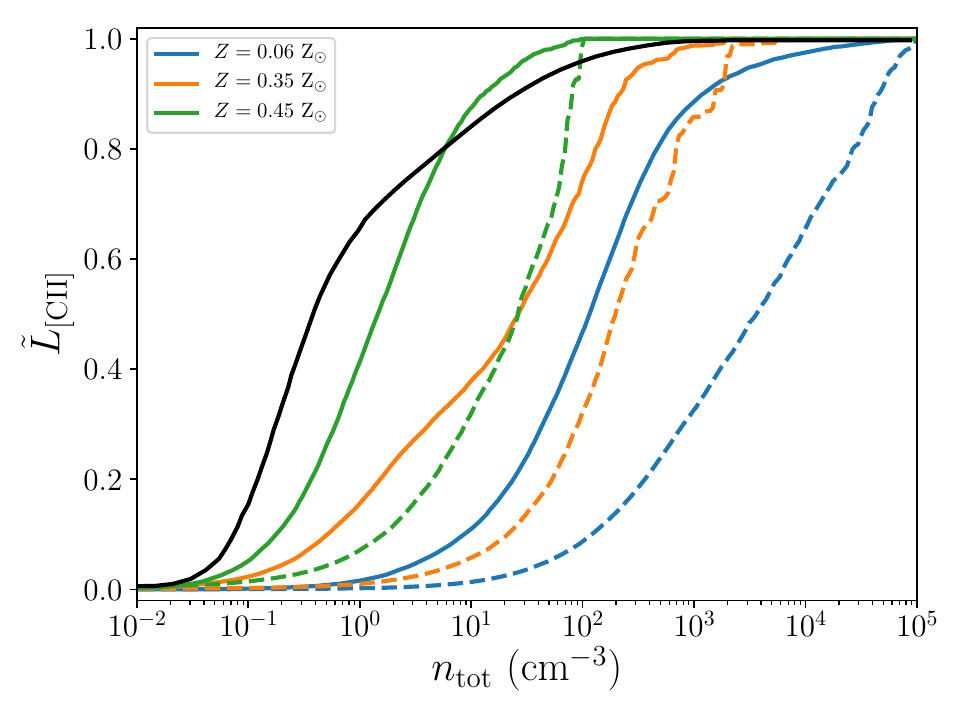}\\
\includegraphics[width=\columnwidth]{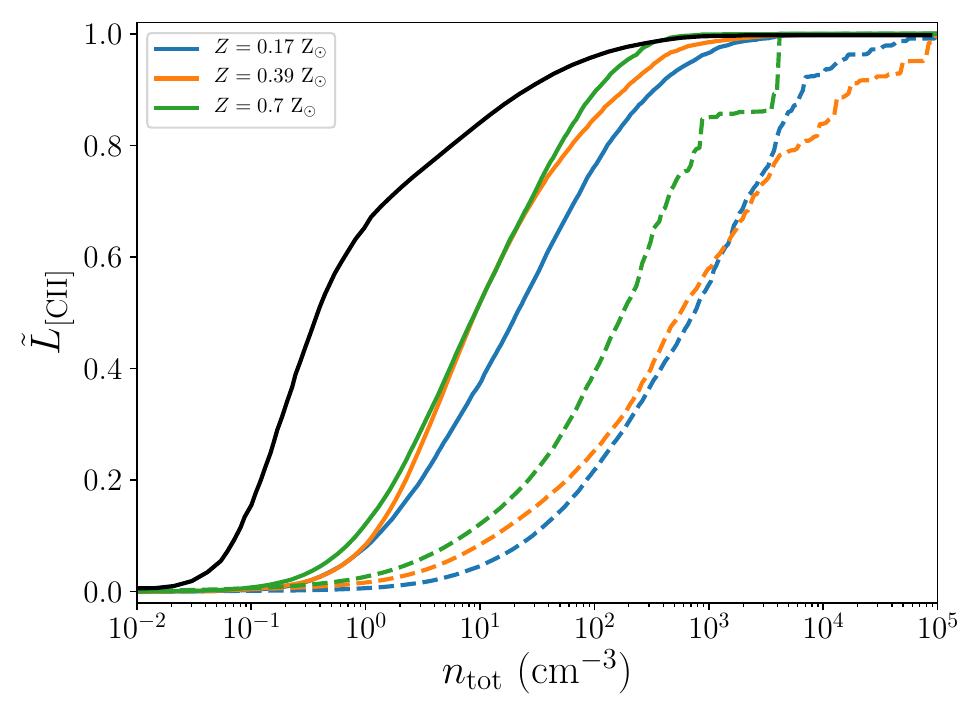}
\caption{Redshift evolution of the normalised cumulative [CII] emission as a function of gas number density for H1 (top panel) and H2 (bottom panel), shown as the coloured solid lines,  for different metallicities. The black line corresponds to the fiducial model by \citet{hu17}. The dashed lines represent instead the normalised cumulative emissivity $L_{\rm [CII]}/M_{\rm gas}$. Our results for H1 look vaguely similar to those by \citet{hu17}, with a significant fraction of the [CII] emission coming from low- and intermediate-density gas, the so-called cold neutral medium, but slightly shifted toward higher densities. For H2, instead, the curves shift toward higher densities, as expected for more massive galaxies where dense gas can survive for longer times before being evacuated by SNe. From the emissivity curves, instead, we see in both panels that the largest contribution comes from dense and colder gas, that accounts for most of the mass, but has a smaller volume filling factor.}
\label{fig:ciicum}
\end{figure}

In Fig.~\ref{fig:ciicum}, we show the normalised cumulative [CII] emission from our galaxies as a function of $Z$($z$), and compare it with isolated dwarf simulations by \citet{hu17}. Similarly to \citet{hu17}, our results suggest that a significant fraction of the total [CII] emission comes from cold neutral gas at low and intermediate densities, and only a small fraction from the high density typical of photo-dissociation regions (PDRs). This is also consistent with the observational results by \citet{fahrion17}, where they observe 46 per cent of the [CII] emission produced in the neutral medium (where the neutral hydrogen emission dominates), and only 9 per cent coming from the cold neutral medium. However, we notice that, unlike \citet{hu17}, the cosmological evolution we considered here results in large variations of the emission distribution, that reflect the evolution of the gas in the galaxies with metallicity. In particular, we can follow our target galaxies throughout the entire cosmic history, self-consistently modelling gas inflows from large scales and galaxy mergers, that affect in a non-trivial way the ISM evolution. By looking at the normalised cumulative emissivity $L_{\rm [CII]}/M_{\rm gas}$, we notice a completely different trend, where most of the emissivity is produced by dense gas, that accounts for a small fraction of the mass only. This suggests that the dense gas, that is typically closer to stellar sources and for which collisions are more frequent, is more effectively excited than its low-density and warmer counterpart, producing a stronger emissivity, but also that its small contribution to the total mass suppresses the total luminosity produced, which is instead dominated by the volume-filling lower density gas. These trends are clearly visible in both our runs, but with some differences. While in H1 both the emission and emissivity distributions evolve significantly with metallicity, shifting to lower density, in H2 the variation is modest, with the two profiles biased toward moderately higher densities. The origin of this difference is two-folds. First, as the metallicity increases, metal cooling becomes more efficient, hence [CII] emission becomes stronger, even at lower densities, as can be easily seen in the top panel for H1 (and also in H2, despite to a smaller degree). Secondly, a more massive halo (like H2) is able to retain dense gas after SN events, whereas smaller haloes like H1 see their dense gas completely evacuated after each SN event.Our results are also consistent with those obtained by \citet{franek18} on synthetic maps of molecular clouds, further reinforcing our conclusions.

\section{Conclusions}
We have presented here two zoom-in cosmological simulations of dwarf galaxies including a self-consistent non-equilibrium metal network \citep{capelo18}. Compared to previous studies by \citet{lupi19a} and \cite{lupi19a}, we have improved the sub-grid models employed, that now include young stellar winds and an updated SF prescription independent of the Mach number. Thanks to the state-of-the-art sub-grid modelling employed, we have been able to well reproduce the main observed properties of local dwarf galaxies. The very high resolution achieved and the ability to directly follow the C$^+$ abundance in the simulation have enabled us to accurately investigate the emergence of the correlation between the [CII] emission and the star formation rate, both spatially resolved and integrated over the entire galaxy, throughout the entire cosmic evolution of the galaxy. Our findings can be summarised as following: 
\begin{enumerate}
    \item the local correlations observed by \citet{delooze14} and \citet{herreracamus15} are very well recovered in our simulation, both in slope and normalisation. Since our data is outside the currently observed range, higher resolution and deeper observations are necessary to
    assess the accuracy of our model in reproducing the [CII]--SFR relation;
    \item at lower metallicity, the correlation seems to evolve in agreement with the predictions by \citet{vallini15}, but this evolution is hard to find in observations because of the large scatter in the \citet{delooze14} data ($\sim 0.42$ dex);
    \item the [CII] emission distribution in dwarfs suggests that most of the emission comes from the CNM ($n_{\rm gas}\lesssim 100~\rm cm^{-3}$), and only a small fraction from the higher densities typical of PDRs, in agreement with the simulation results by \citet{hu17} and \citet{franek18}, and SOFIA observations by \citet{fahrion17}. This is due to the large volume filling factor of the low-density gas, as demonstrated by the normalised cumulative emissivity, that is dominated by the cold and dense gas (likely in PDRs) where most of the mass is confined but the occupied volume is small.
\end{enumerate}
Of course, the intrinsic limitations of this study, like the missing additional processes like HII regions around stars \citep{hu17,hopkins18a}, cosmic rays, etc., the inability to properly resolve the PDRs, and the missing galaxy population statistics (we only considered two galaxies due to the computational cost of the runs), does not allow us to give a more conclusive answer on the  metallicity dependence of the [CII]--SFR correlation. Nevertheless, the agreement found in the evolution of two completely different galaxies is a hint that our conclusion could be real. This study is only meant to represent a first attempt to address this important question and the related consequences about the significance of the [CII] as a SFR tracer, and a more detailed exploration is definitely necessary.

\section*{Acknowledgements}
We thank the anonymous referee for useful comments that helped improving the quality of the manuscript.
The authors thank Rodrigo Herrera-Camus and Livia Vallini for fruitful discussions.
AL acknowledges support from the European Research Council No. 740120 `INTERSTELLAR'. SB is financially supported by Fondecyt Iniciaci\'on (project code 11170268) and Conicyt Apoyo a la Formaci\'on de Redes Internacionales project number REDI170093. SB also acknowledges support by BASAL Centro de Astrofisica y Tecnologias Afines (CATA) AFB-17002. The authors acknowledge the UdeC Kultrun Astronomy Hybrid Cluster (projects Conicyt Programa de Astronomia FondoQuimal QUIMAL170001, Conicyt PIA ACT172033,  and Fondecyt Iniciaci\'on 11170268) for providing HPC resources that have contributed to the research results reported in this paper.




\bibliographystyle{mnras}
\bibliography{Biblio} 


\appendix
\section{Validation of the chemical model for cosmological simulations: the galaxy mass dependence of the [CII]--SFR and [OI]--SFR correlations}
\label{app:oirelation}
\begin{figure}
\centering
\includegraphics[width=\columnwidth]{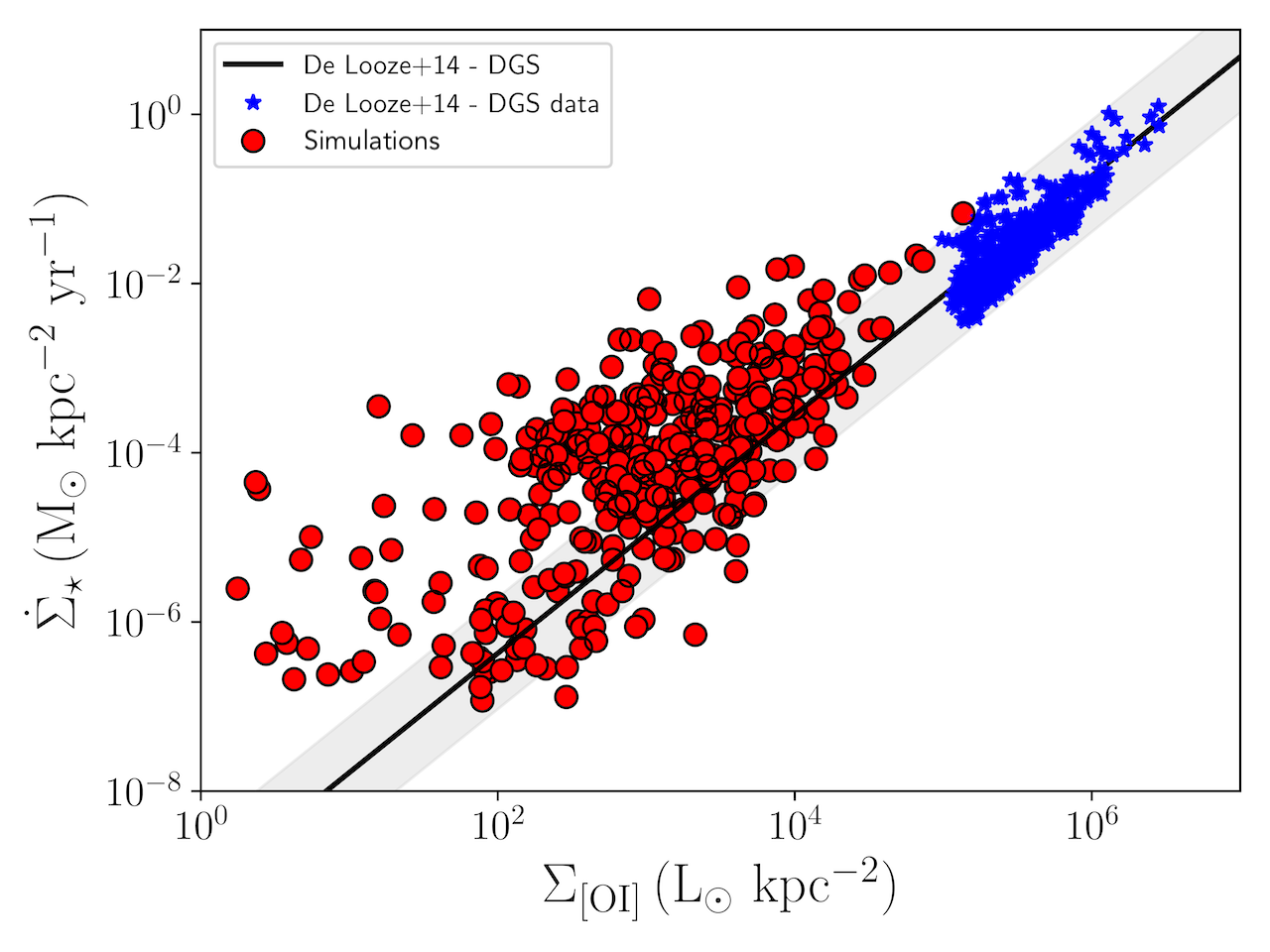}
\caption{Spatially-resolved [OI]--SFR correlation in the simulations, compared with the best-fit by \citet{delooze14}, shown as a black solid line, where the grey shaded area corresponds to the $3\sigma$ uncertainty. Our simulations (red dots), binned in square patches 500~pc wide, appear to be offset with respect to the observed correlation. This is likely due to the lower abundance of moderately dense regions of gas in the galaxies, consequence of the shallow potential well and the strong impact of SN feedback. In H2, where the galaxy is more massive, the data start to approach the observed correlation, although the agreement is not recovered yet.}
\label{fig:OIres_dwarf}
\end{figure}

\begin{figure}
\centering
\includegraphics[width=\columnwidth]{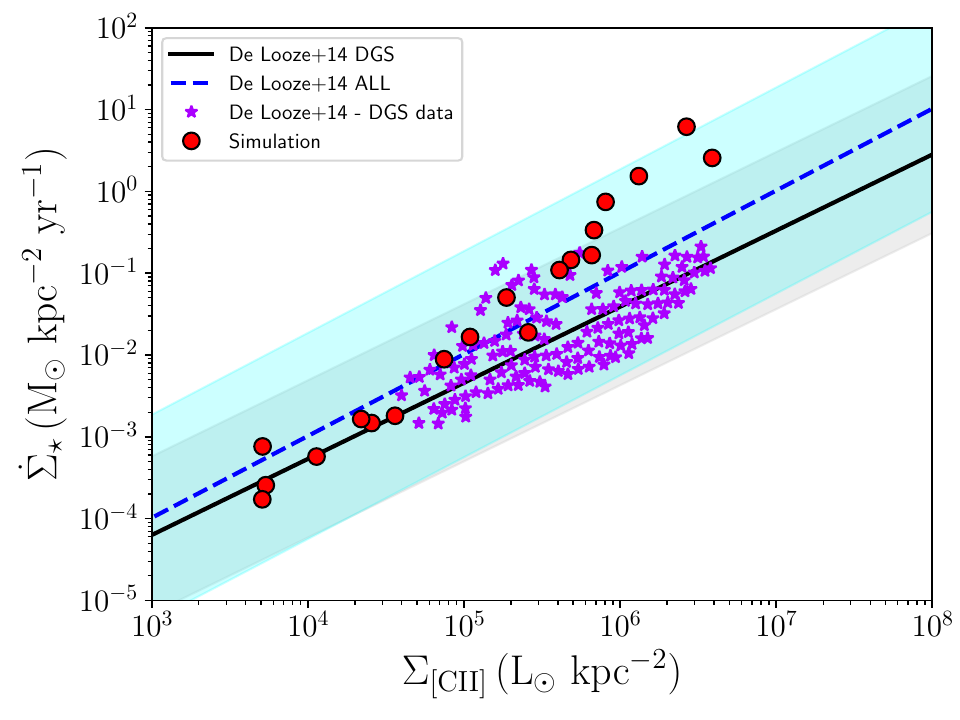}
\includegraphics[width=\columnwidth]{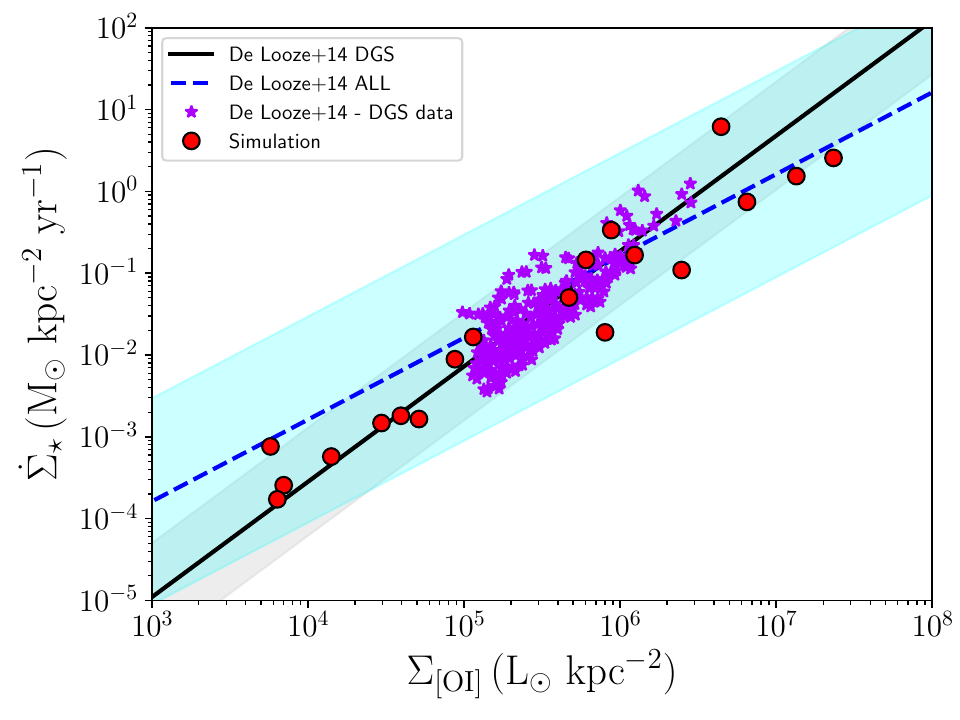}
\caption{[CII]--SFR (top panel) and [OI]--SFR (bottom panel) correlations for the Milky Way-like halo. The purple stars correspond to the DGS galaxies by \citet{delooze14}, with the black solid line (and the grey shaded area) representing the best-fit (and the $3\sigma$ scatter) over the DGS sample, and the blue dashed line (and the cyan shaded area) the best-fit to the entire literature sample by \citet{delooze14}. Our simulation data, shown as red dots, exhibits a very good agreement with observations in both panels, because of the presence of dense regions in the galaxy where [OI] emission is much stronger. While the [OI] exhibits a consistent trend across the entire luminosity range, the [CII] counterpart shows a deficit at very-high luminosities. This deficit is due to the presence of very-high density gas in the central region of the galaxy, where the C$^+$--C transition is efficient, and the [CII] emission is suppressed.}
\label{fig:MWcorrel}
\end{figure}

As in \citet{capelo18}, the chemical network employed here includes both C and O atoms, the main elements responsible for radiative cooling in the ISM below $10^4$~K. In their isolated galaxy simulations, \citet{capelo18} showed that both the [CII]--SFR and the [OI]--SFR correlations could be reproduced at the same time in sub-L* galaxies with sub-solar metallicity. However, in a cosmological context, this is not necessarily guaranteed. Indeed, in our dwarf galaxy simulations, while the [CII]--SFR correlation exhibits good agreement with the extrapolation of the observed correlation, the [OI]--SFR is moderately offset (see Fig.~\ref{fig:OIres_dwarf}). From a physical point of view, this behaviour can be explained by considering that, while [CII] is also excited in the diffuse neutral medium, the [OI] emission is dominated by denser gas closer to SF regions, which in our simulations are rare and short-lived due to SN feedback. 
To test this idea, and exclude a possible numerical origin of this result (due to the resolution or the sub-grid modelling), we have performed a cosmological simulation of a more massive galaxy with the same sub-grid models of the fiducial run. The initial conditions are taken from the FIRE2 simulation suite \citep{hopkins18a}, and correspond to a Milky Way-like halo (namely the {\it m12i} case). To limit the computational cost of the run, we employed the low-resolution version, where the mass resolution $m_{\rm gas}=5.6\times 10^4\,\msun$.
We evolve the galaxy down to $z=2$, when the metallicity is already comparable to that of our dwarf galaxy at $z=0$, and we repeat the analysis performed in Section \ref{sec:ciiphase}. The [CII]--SFR and [OI]--SFR correlations are reported in the top and bottom panels of Fig.~\ref{fig:MWcorrel}, respectively, at $z=2$. In this case, to better highlight the trend of the distribution and limit the number of data points in the plot, the simulation data has been binned in annuli 500~pc wide, with the results shown as red dots. Because of the larger mass of the galaxy compared to our dwarf, SN feedback becomes less effective, and the gas can accumulate in the galaxy and more easily reach the typical densities where [OI] emission is powered.

Consistently with many observations of high-redshift galaxies, we also observe a [CII] deficit at very high luminosities, obviously not reproduced by the best-fit relation by \citet{delooze14}. For what concerns the [OI] emission, instead, the agreement is very good across the entire luminosity interval. Although not reported, we also confirm the presence of an evolution of the integrated [CII]--SFR correlation with metallicity, similarly to the dwarf galaxy case.
Concluding, the chemical network employed in this study works well also in cosmological simulations and is able to reproduce both the [CII]--SFR and [OI]--SFR correlations when the conditions to trigger efficient emission are met. Hence, the poorer agreement found in our dwarf galaxies is likely due to the lack of denser gas where the [OI] line is more efficiently excited, gas that is quickly dispersed by SN feedback.
As a further comparison, we computed the [OI]/[CII] ratio for both dwarf and the Milky Way-like galaxies, finding an almost constant value across the entire SFR surface density interval sampled, as found in the Small Magellanic Cloud by \citet{jameson18}. However, the ratio in our simulations is about unity, in better agreement with the results by \citet{herreracamus15} for low-metallicity systems. 

\bsp	
\label{lastpage}
\end{document}